%Paper: hep-th/9407183
%From: Peter West <pwest@mth.kcl.ac.uk>
%Date: Wed, 27 Jul 94 14:00:21 +0100
%Date (revised): Fri, 7 Oct 94 16:51:38 +0100

\input phyzzx

\Pubnum{ \vbox{  \hbox{KCL-TH-94-13} } }
\pubtype{}
\date{July 15, 1994}

\titlepage

\title{Free Fermions and Extended Conformal Algebras }

\author{P.S. Howe$^1$, \quad G. Papadopoulos$^2$\foot{
Address from 1st October:  DAMTP, University of Cambridge, England},
\quad P.C.
West$^1$}
\address{1.\quad Department of Mathematics,
King's College London, England
\break
2.\quad II. Institute for Theoretical Physics,
 University of Hamburg,  Germany
\break
and
\break
Blackett Laboratory, Imperial College, England}

\abstract{A class of algebras is constructed using free fermions
and the
invariant antisymmetric tensors associated with irreducible
holonomy groups.}
\endpage
%%%%%%%%%%%%%%%%%%%%%%%%%%%%%%%%%%%%%%%%%%%%%
%%%%%%%%
%%%%%%%%%%%%%%%%%%%%%%%%%%%%%%%%%%%%%%%%%%%%%
%%%%%%%%%%%%%%%%%

\font\mybb=msbm10 at 12pt
\def\bb#1{\hbox{\mybb#1}}
\def\RN {\bb{R}}

\def\a{\alpha}
\def\b{\beta}
\def\c{\gamma}
\def\d{\delta}
\def\l {\lambda}

\REF\witten {E. Witten, Commun. Maths. Phys. {\bf 92} (1982) 455.}

\sequentialequations

%%%%%%%%%%%%%%%%%%%%%%%%%%%%%%%%%%%%%%%%%%%%%
%%%%%%%%%%%%%%%%%%

%\chapter{}

Given a set of free fermions $\{\lambda^a; a=1,\dots,n\}$ in two
dimensions,
one can construct a set of conserved currents
$\{J^{(p)}; p=1,\dots,n\}$ where
$J^{(p)}$ is the normal-ordered product of $p$ fermion fields (with
uncontracted indices).  Thus each current is an antisymmetric tensor
in an
associated n-dimensional vector space $V$ ($\rm{dim} V=n$).  The set
of currents
which are quadratic in the fermion fields ($p=2$) form an $SO(n)$
Kac-Moody
current.  As such, it is natural to ask whether there are other current
algebras
constructed by taking higher order polynomials of this currents.
Indeed it was observed [1] that one could generate the $N=1$
superconformal algebra by  constructing the supercharge from
the three fermion current contracted
with the structure constant of a Lie algebra.

In this note we will show that certain subalgebras of this larger
algebra may be
extracted using antisymmetric tensors of $V$ that are invariant
under of a
subgroup $G$ of $O(n)$ acting on $V$ with some representation $R$.
Invariant
forms can exist in every irreducible representation of a compact
Lie group.  To
be more precise, we shall use the invariant forms which occur in
the fundamental
representation of irreducible holonomy groups of non-symmetric
Riemannian
manifolds.  We will also use the invariant forms of the adjoint
representation of these groups.  Typically we find algebras with
additional spin $3/2$ and spin $2$ generators.
For $SO(n)$ and $SU(n/2)$
higher spin currents arise for large enough $n$.
In these cases, one
does not generate a closed algebra using only the invariant
tensors (and
$T$) although it might be possible to achieve closure by including a
finite number of additional currents.

We recall that the possible irreducible holonomy groups of
n-dimensional
non-symmetric Riemannian manifolds are given by Berger's list [2]:
$SO(n)$,
$U(n/2)$, $SU(n/2)$, $Sp(n/4)$,
$Sp(n/4)\cdot Sp(1)$ and two
exceptional cases,
$G_2$ ($n=7$) and Spin(7) (n=8).  The corresponding invariant
forms are:
the $\epsilon$-tensor ($SO(n)$); the Kahler 2-form ($U(n/2)$),
the
holomorphic $\epsilon$-tensor ($SU(n/2)$), the three Kahler
2-forms
($Sp(n/4)$), and for  $Sp(n/4)\cdot Sp(1)$, a four-form which is
locally the sum of
the wedge product of each of the three local Kahler forms with
itself.  In $G_2$, there is a 3-form and its 4-form dual, and  in
$Spin(7)$ a self-dual 4-form.  It has been noted [3] that
(two-dimensional) supersymmetric sigma models with these manifolds as
target spaces have additional symmetries associated with these form;
and that the corresponding currents belong, at least classically, to
extended superconformal algebras which close in a non-linear fashion,
i.e. they
are of the W-symmetry type.
The $SU(3)$, $G_2$ and $Spin(7)$ cases are
relevant in string theory compactifications to $4,3$ and $2$
dimensions respectively (see for example [5]).

Let us now turn to a free fermion theory with $n$ fermions
$\{\lambda^a\}$.  In the general case, we can introduce the currents
$$
J_{a_1\cdots a_p}=:\lambda_{a_1}\cdots \lambda_{a_p}:
\eqno (1)
$$

The OPE of two such currents (with $q\leq p$) is
$$
J_{a_1\cdots a_p}(z) J_{b_1\cdots b_q}(w) = \sum^q_{m=1}
{(-1)^{pm+{m(m-1) \over 2}} \over (z-w)^m}
\delta _{[a_1 \ldots a_m}^{[b_1 \ldots b_m}
\lambda_{ a_{m+1} \ldots a_p]} (z) \lambda^{b_{m+1} \ldots b_q]} (w)
\eqno (2)
$$
The antisymmetry symbols in the above equation are with weight
one. Clearly, this O.P.E algebra closes in the sense that
all  the
terms on the right-hand-side can be arranged into products of $J$'s
and their derivatives.

To extract the sub-algebras of interest we shall
contract some of the
above J's with the invariant tensors of the holonomy
groups $G$ which
we have listed above.  Let
$$ g=g_{a_1\dots a_k} e^{a_1}\wedge \dots\wedge e^{a_k}
$$
be a (constant)  k-form of the vector space $V$ and
$\{e^a\}$, $a=1,\dots,
\rm{dim} V$ a basis of
$V$.  The  currents that we will study are
$$
X={1\over k!} g_{a_1\dots a_k} :\lambda_{a_1}\cdots
\lambda_{a_k}:
\eqno(3)
$$
All these currents are primary with conformal spin
${k\over 2}$ with respect to
the energy momentum
$$
T= {1\over 2}\l^a\partial\l_a
$$
of the free fermions, i.e. the O.P.E of $T$ and $X$ is
$$
T(z) X(w) = {\partial X\over z-w}+{k\over 2} {X\over (z-w)^2}.
\eqno(4)$$

\section{\bf $G=SO(4)$}

We define
$$
X={1\over 4!} \epsilon_{abcd} :\l^a \l^b\l^c\l^d:
\eqno(5)
$$
we then find
$$
X(z)X(w)= {1\over {(z-w)}^4} + {2T \over {(z-w)}^2}
+ {\partial T \over (z-w)}
\eqno(6)
$$
The O.P.E of $T$ with $X$ closes as in equation (4).
Thus, we have an algebra with two spin 2 currents.  It might
be thought that
this algebra should factorise, but it does not as one may
verify by considering
all possible linear combinations of $X$ and $T$.

If one goes to higher $n$,  one finds that the OPE algebra does not
close on $T$ and the spin $n/2$ current
$$
X={1\over n!} \epsilon_{abcde\ldots}
:\l^a \l^b\l^c\l^d\l^e\ldots :
\eqno(7)
$$
Indeed, if one considers the OPE of $X$ with itself, the
two most singular terms, are
$${{(-1)}^{n  (3n-1) \over 2} \over {(z-w)}^n} -
{{(-1)}^{n(3n+1)\over 2} \over {(z-w)}^{n-2}} .
{1 \over 2}. \lambda^{ab}(z) \lambda_{ab}(w).\eqno(8) $$

Up to a sign we can identify the coefficient of the latter
term as $2 T^2$ .  However, were the algebra to close on $X$ and $T$
alone then the coefficient of the above $T^2$ term is fixed [4]
, by conformal invariance, to be $2.{5s+1 \over 22+5c}$
in magnitude,  where, for us,
$c=s={n\over 2}$.  Clearly, the above O.P.E. can not close on $X$ and
$T$ alone.  Explicit calculation of the full O.P.E. for SO(5) bears
out this conclusion.  This does not mean that the O.P.E. does not
close if one includes more currents or a different choice for
energy-momentum tensor.

\section{\bf $G=SU(3)$}

In $SU(3)$ we have $n=6$ $\l$'s and replace them by 3 complex
ones, i.e.
$\l^a\rightarrow \{\l^\a, \bar{\l}^\a\}$, $\a=1,2,3$.
There is complex spin
$3/2$ current
$$
G={1\over 3!} \epsilon_{\a\b\c} :\l^\a\l^\b\l^\c:
\eqno(9)
$$
and its conjugate
$$
\bar{G}={1\over 3!} \epsilon_{\a\b\c}
:\bar{\l}^\a\bar{\l}^\b\bar{\l}^\c:
\eqno(10)
$$
One finds
$$
G(z) \bar{G}(w)= {1\over (z - w)^3} - {J\over(z-w)^2}
+ {\{- {1\over 2} \partial J + T^\prime\} \over (z-w)}
\eqno(11)
$$
where
$$
J=:\l^\a \bar{\l}^\a:\ \ T^\prime = T
+ {1\over 2} : \lambda^\beta \bar \lambda_\beta \lambda^\gamma \bar
\lambda \gamma :
\eqno(12)
$$
While $J$ is a $U(1)$ current, one can show by re-normal
ordering the $\l^4$ term that $T^\prime  = {1\over 2} : J^2 :$ and so
is the Sugawara energy-momentum tensor associated with the the $U(1)$
current.
Evaluating the other O.P.E's and, making a suitable rescaling, one
finds that
$T^{\prime} , G, \bar G$ and
$J$ define an $N=2$ superconformal algebra with central charge $c=1$.
The field $T - {T^\prime \over 3}$ commutes with this algebra. This algebra
occurs as a subalgebra of the extended $N=2$ superconformal algebra which
arises in the context of Calabi-Yau compactifications [6].

For the $SU(n)$ case, we have the spin ${n\over 2}$ current
$$
X={1\over n!} \epsilon_{\a\b\c\d\ldots} :\l^\a\l^\b\l^\c \l^\d\ldots:
\eqno(13)
$$
and its complex conjugate
$$
\bar{X}={1\over n!} \epsilon_{\a\b\c\d\ldots}
:\bar{\l}^\a\bar{\l}^\b\bar{\l}^\c\bar{\l}^\d\ldots:
\eqno(14)
$$
The OPE of $X$ with its complex conjugate does not close on $T$
for the reasons given earlier, however one may hope that it will
close using Sugawara-like currents. This possibility remains under study.

\section{\bf $Sp(k)\cdot Sp(1)$}

This is the holonomy group of quaternionic Kahler manifolds
in which case
there are three (locally defined) complex structures $I_r$ that
obey the algebra of
imaginary unit quaternions
($I_rI_s=-\delta_{rs}+\sum_t \epsilon_{rst} I_t$) and
three associated Kahler forms
$\omega_r$.  From these one can define an $Sp(k)\cdot Sp(1)$
invariant 4-form
$$
\Omega=\sum_{r=1}^3 \omega_r\wedge \omega_r
\eqno(15)
$$
The associated spin 2 current current is
$$
X={1\over 4!}\Omega_{abcd} :\l^a \l^b\l^c\l^d
\eqno(16)
$$
One finds that the OPE algebra is
$$\eqalign{
X(z)X(w)&= {1\over 4!} {n(n+2)\over (z-w)^4}
+{(n+2)\over 3} {T(w) \over(z-w)^2}
+ {(n+2)\over 6}{\partial T (w)\over z-w}
 -{n-4\over 6}{\partial X(w)\over z-w}
\cr
-&{n-4\over 3} {X(w)\over (z-w)^2},}
\eqno(17)
$$
where $n=4k$.

\section{\bf $G_2$ and Spin(7)}

The final examples are provided by the exceptional
holonomy groups.  In $G_2$
there is a 3-form ($e^a$ a basis in $\RN^7$)
$$
\phi=f_{abc} e^a\wedge e^b \wedge e^c
\eqno(18)
$$
and a 4-form which is the dual of $\phi$
$$
^*\phi=f_{abcd}e^a\wedge e^b \wedge e^c\wedge e^d
\eqno(19)
$$
Using these we can form the currents
$$
G={1\over 3!} f_{abc}:\l^a\l^b\l^c:
\eqno(20)
$$
and
$$
X={1\over 4!}f_{abcd}:\l^a\l^b\l^c\l^d:
\eqno(21)
$$
The OPE algebra is
$$
\eqalign{G(z)G(w)& =  {7\over {(z-w)}^3} + {6T\over {(z-w)}} -
{6 X \over {(z-w)}}\cr
 G(z) X(w) &=-{6 G(y)\over (z-w)^2}-{2\partial G\over z-w}\cr
 X(z)X(w) &= {7\over (z-w)^4 } +{8T\over ( z-w)^2}
+{4\partial T\over z-w} - {6 X\over (z-w)^2}
-{3\partial X\over z-w}.\cr}
\eqno(22)
$$

In the Spin(7) case ($n=8$), one has an invariant 4-form
which is closely
related to the forms of $G_2$;
$$
\phi=g_{abcd} :\l^a\l^b\l^c\l^d:
\eqno(23)
$$
with $g_{0abc}=f_{abc}$ (for $a,b,c=1,\dots,7$) and
$g_{abcd}=f_{abcd}$ (for
$a,b,c,d=1,\dots,7$).  The associated spin 2 current is
$$
X={1\over 4!} g_{abcd}:\l^a\l^b\l^c\l^d:
\eqno(24)
$$
The O.P.E is
$$\eqalign{
X(z)X(w)= {14\over (z-w)^4}} +{14 T\over (z-w)^2}
+ {7\partial T\over z-w}- {12 X\over (z-w)^2}
-{6\partial X\over z-w}\eqno(25)
$$
\vskip 1cm
{\bf Note Added}

This paper was based on work carried out a number of years ago, we were
motivated to publish it by the recent appearance of the paper
"Superstrings and Manifolds of
Exceptional Holonomy ". by S. Shatashvili and C. Vafa in which
the $G_2$ and spin(7) examples are discussed in a free superfield realisation.

\vskip 1cm

{\bf References}

\item {1} P. di Vecchia, V. Knizhnik, J. Peterson, P. Rossi; Nucl Phys
B253 (1985) 157.
\item {2} M. Berger, Bull. Soc. Math. France 83 (1953) 279
\item {3} P. S. Howe and G.Papadopoulos, Phys.lett. 263B (1991)
230. \hfil \break
 P. S. Howe and G.Papadopoulos, Phys.lett .267B (1991) 362.
\hfil \break
 P. S. Howe and G.Papadopoulos, Comm Math Phys 151 (1993) 467.
\item {4} A.B. Zamolodchikov, Theor. Math. Phys 65 (1989) 1205.
\item {5} P. Candelas, G. Horowitz, A. Strominger and E. Witten,
Nucl. Phys. B258 (1985) 46.  \hfil \break
 C. M. Hull, " Superstring Compactifications with Torsion and space-time
Supersymmetry", Published in Turin conference on "Superunification"
(1985) 347.
\item {6} S. Odake, Mod. Phys. Lett A4 (1989) 557.
\end